\documentclass[letterpaper]{article} 
\usepackage{aaai2026}  
\usepackage{times}  
\usepackage{helvet}  
\usepackage{courier}  
\usepackage[hyphens]{url}  
\usepackage{graphicx} 
\usepackage{natbib}  
\usepackage{caption} 
\usepackage{tabularray}
\usepackage[table]{xcolor}
\usepackage{booktabs}
\usepackage{svg}
\usepackage{amsmath,amssymb}
\usepackage{multirow}

\usepackage{algorithm}
\usepackage{algorithmic}

\usepackage{newfloat}
\usepackage{listings}
\DeclareCaptionStyle{ruled}{labelfont=normalfont,labelsep=colon,strut=off} 
\floatstyle{ruled}
\newfloat{listing}{tb}{lst}{}
\floatname{listing}{Listing}
\title{Veli: Unsupervised Method and Unified Benchmark for Low-Cost Air Quality Sensor Correction}
\author{
    Yahia Dalbah\textsuperscript{\rm 1},
    Marcel Worring\textsuperscript{\rm 1},
    Yen-Chia Hsu\textsuperscript{\rm 1}
}
\affiliations{
    \textsuperscript{\rm 1}University of Amsterdam\\

    y.i.r.dalbah@uva.nl,
    m.worring@uva.nl,
    y.c.hsu@uva.nl
}

\usepackage{bibentry}

\begin{document}
\maketitle

\begin{abstract}
Urban air pollution is a major health crisis causing millions of premature deaths annually, underscoring the urgent need for accurate and scalable monitoring of air quality (AQ). 
While low-cost sensors (LCS) offer a scalable alternative to expensive reference-grade stations, their readings are affected by drift, calibration errors, and environmental interference. 
To address these challenges, we introduce \textbf{Veli} (Reference-free \textbf{V}ariational \textbf{E}stimation via \textbf{L}atent \textbf{I}nference), an unsupervised Bayesian model that leverages variational inference to correct LCS readings without requiring co-location with reference stations, eliminating a major deployment barrier.
Specifically, Veli constructs a disentangled representation of the LCS readings, effectively separating the true pollutant reading from the sensor noise. To build our model and address the lack of standardized benchmarks in AQ monitoring, we also introduce the Air Quality Sensor Data Repository (AQ-SDR).
AQ-SDR is the largest AQ sensor benchmark to date, with readings from 23,737 LCS and reference stations across multiple regions. Veli demonstrates strong generalization across both in-distribution and out-of-distribution settings, effectively handling sensor drift and erratic sensor behavior. Code for model and dataset will be made public when this paper is published. The appendices are available in the extended version.


\end{abstract}

\begin{links}
    \link{Code}{https://github.com/YahiDar/Veli}
    \link{Datasets}{https://github.com/YahiDar/AQ-SDR}
\end{links}

\section{Introduction}
\label{sec:introduction}

The World Health Organization (WHO) estimated that over 90\% of the world's population breathes air that contains pollutants above WHO guideline levels \cite{WHO2025HealthRisks}. 
These pollutants are known to cause respiratory and cardiovascular diseases, and are present in high concentrations in urban areas \cite{Zhang2025PM2_5CVD}. 
To meet WHO air quality standards, real-time air quality (AQ) monitoring is crucial. 

Municipalities and environmental agencies rely on well-maintained, expensive monitoring stations to report pollution at the district level.
The high cost of buying, installing, and maintaining these stations makes it infeasible to achieve the spatial coverage needed to capture microclimates affecting citizens.
Consequently, numerous initiatives have emerged to scale up the spatial coverage of AQ sensing by using low-cost sensors (LCS). 
In contrast to expensive monitoring stations, LCS are affordable and accessible to the average citizen, making them suitable for crowdsourcing projects. 
However, LCS produce raw data that are inaccurate, noisy, and often unreliable, making it difficult to use their readings to make informed decisions. 

\begin{figure}[t]
  \centering
  \includegraphics[width=0.95\columnwidth]{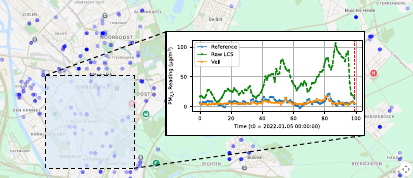} 
  \caption{A snapshot from the AQ-SDR dashboard of sensors in the city of Utrecht in the Netherlands. The query area shows the results of applying our method, Veli, on hourly noisy readings from deployed LCS over four days.}
  \label{fig:hourly-avgs-utrecht}
\end{figure}
To use LCS to increase the spatial coverage of AQ monitoring, reliable methods for correcting their erratic readings are necessary.
Many pre-deployment calibration methods exist for LCS \cite{Delaine2019Calibration,Hagan2018Sensors,Maag2016PreDeployment}.
However, dense deployment of LCS would require recurrent manual recalibration to prevent issues like sensor drift.
To eliminate the need for manual recalibration, numerous works have explored numerical approaches for post-deployment data correction\footnote{To avoid confusion, we use the term `correction' for all numerical/algorithmic approaches to data processing, and distinguish it from instrument calibration of the devices.}.
LCS data correction methods often rely on high-cost reference stations as the ground truth to train supervised machine learning models. 
A fundamental limitation of these models is 
their reliance on the co-location of LCS with high-cost stations to collect synchronized data pairs for training, which undermines the core objective of using LCS as an affordable option to increase spatial coverage \cite{survey2}.
Moreover, these data correction models are typically trained over a short period of time (often a few months), making them unreliable for long-term applications due to sensor drift and seasonal variations. 
Another significant but largely overlooked limitation is that these models often fail to account for real-world operational issues.
For instance, deployed LCS exhibit significant bias and drift, and can experience periods of data or connectivity loss, causing their uncorrected readings to mislead end-users and public health analysts \cite{Concas2021LowCost}. 
Lastly, previous studies do not use a standard benchmark or dataset for model evaluation. 
The lack of a common benchmark hinders reliable evaluation, as reported metrics often lack the context to compare different methods effectively.

To address these challenges, we introduce \textbf{Veli} (reference-free \textbf{V}ariational \textbf{E}stimation via \textbf{L}atent \textbf{I}nference), an unsupervised post-deployment LCS correction model. 
To develop and test our model, we built a standardized benchmark for AQ research, the Air Quality Sensor Data Repository (AQ-SDR). 
Our work makes three primary contributions:

\begin{itemize}
    \item We propose a novel reference-free method for unsupervised data correction, eliminating the need for co-location with high-cost reference stations.
    \item We release the largest public benchmark for AQ monitoring, containing 23,737 sensors across diverse regions and pollution levels. This benchmark contains common sensor errors and operational failures, providing a resource suitable for modeling practical LCS deployment.
    \item We validate the model's real-world effectiveness and demonstrate its robustness and generalizability in both in-distribution and out-of-distribution settings.
\end{itemize}

\section{Related Work}
\label{sec:relatedwork}

We categorize prior work into two groups: methods that rely on expensive, well-maintained reference stations for training (reference-based methods) and methods that do not use reference stations for training, and only use them for model evaluation (reference-free methods). In this work, we use the terms reference-free and unsupervised interchangeably.
In the absence of established reference-free methods, we contextualize our contribution through a review of current reference-based approaches.


\subsection{Reference-based Methods}
\label{subsection:referencebasedmethods}
\subsubsection{Reference-based Correction Methods}
Reference-based correction methods use reference stations to correct inaccurate LCS readings. 
Given two sets, \(X_{\mathrm{LCS}}\) and \(Y_{\mathrm{ref}}\), synchronized in time, a model \(\textbf{M}\) is trained to minimize the deviation between \(Y_{\mathrm{ref}}\) and the mapping ${\mathbf{M}}(X_{\mathrm{LCS}})$ (e.g., using mean squared error). We assume by default that all reference ground truth data originate from accurate, well-maintained instruments.
Reference-based correction methods are split into pre-deployment or post-deployment methods, depending on when the correction occurs.

A major limitation of pre-deployment reference-based correction methods is the need to co-locate target LCS units next to a reference station for an extended period to collect calibration data, making the deployment of large LCS networks impractical. 
Moreover, shorter co-location intervals yield models that poorly capture temporal variations such as seasonal changes. 
Lastly, this initial calibration does not account for long-term sensor drift, necessitating periodic recalibration. The logistical challenges of recalibrating deployed sensors mean that long-term drift often goes uncorrected in many devices.
Most early studies adopted simple linear models in the pre-deployment context (e.g., ordinary least squares regression). We refer readers to \cite{Concas2021LowCost,survey2} for a comprehensive review of these methods. 

The complexity of LCS errors has recently led to increased interest in non-linear post-deployment correction methods.
These methods address the limitations of the traditional design paradigm, which relies on synchronized and co-located LCS-reference pairs, similar to \cite{tesla}. 
For instance, \cite{Cheng2019ICT} addressed post-deployment correction using unsynchronized calibration transfer, a technique for in-field calibration via co-location with a reference station.
This co-located LCS then serves as an anchor point, providing ground truth for other sensors with no co-located references in the network.
While this method reports promising results, it was tested on only seven LCS during a ten-month period. 
Moreover, the LCS units were deployed in controlled settings, avoiding real-world issues such as missing data and extreme fluctuations.
\cite{Wang2023CaliFormer} proposed \textit{CaliFormer}, a hybrid reference-based approach combining unsupervised reconstruction with supervised fine-tuning. 
The model is initially trained to reconstruct the LCS data in an unsupervised manner, and then fine-tuned to correct the results using ground-truth data from the reference stations.

In addition to direct correction of readings, some works have used historical data from reference stations as prior knowledge for LCS correction.
Both `RHC' \cite{rch2020} and the Maximal Correlation Model~\cite{maximizingcorrelation} leverage historical reference data to align LCS and reference readings' distributions.
A significant limitation is that both approaches were evaluated on short time frames, restricting their applicability for long-term deployments.

\subsubsection{Reference-based Interpolation Methods}

A different line of work bypasses LCS correction altogether, creating high-resolution AQ maps by interpolating data directly from a network of reference stations.
Both MapTransfer \cite{Cheng2020MapTransfer} and AirRadar \cite{Wang2025AirRadar} interpolate readings from high-cost stations to generate denser pollution maps. Despite their ability to produce high-resolution AQ maps, these approaches depend on reference stations with a sparse deployment across a region, which limits their ability to capture microclimate variations.



\subsection{AQ Benchmarks}
\label{subsection:aqdatasets}

In Table \ref{tab:dataset_comparison}, we compare previously published datasets and benchmarks that contain LCS data with our new dataset, AQ-SDR. We provide further details on AQ-SDR in Section \ref{subsection:AQ-SDR} and Appendix \ref{appendix:dataset}.
Previous datasets are either limited to small-scale studies on a regional level \cite{Diez2024QUANT}, or cover shorter time periods \cite{Jiao2016CAIRSENSE}. 
While some benchmarks provide aligned LCS and reference station readings \cite{Bi2022,VanPoppel2023SensEURCity}, they do not provide a scale large enough to develop models that can generalize across diverse pollution levels.
Our dataset is designed to serve as a unifying benchmark for LCS modeling and correction methods, capturing a wide range of failure modes, distribution shifts, sensor drift, and pollution levels to reflect real-world LCS behavior.
AQ-SDR is the largest AQ sensor dataset to date, containing data from 23,737 low-cost and reference sensors across multiple global regions, collected over more than six years of deployment.

\begin{table}[t]
  \centering
  \small
  \begin{tabular}{l | c c c}
    \specialrule{1pt}{0pt}{0pt}  
    \multirow{2}{*}{\textbf{Dataset}} & \textbf{\# of} & \textbf{Period} & \textbf{LCS \&}  \\
     & \textbf{Sensors} & \textbf{(months)} & \textbf{Reference}  \\ \hline
    \cite{Jiao2016CAIRSENSE} & 20 & 10 & LCS Only  \\
    \cite{Diez2024QUANT} & 49 & 34 & LCS Only \\
    \cite{VanPoppel2023SensEURCity} & 85 & 12 & Both  \\
    \cite{Bi2022} & 109 & 22 & Both  \\
    \hline
    AQ-SDR & 23737 & 80 & Both  \\ 
    \specialrule{1pt}{0pt}{0pt}   
  \end{tabular}
  \caption{Comparison between our dataset and other published AQ datasets. We disregard small-scale hyperlocal studies and datasets that have fewer than 10 sensors.}
  \label{tab:dataset_comparison}
\end{table}

\section{Reference-Free LCS Correction}

\label{sec:referencefreelcscorrection}

\subsection{Problem Formulation}
\label{subsection:problemformulation}
A key challenge for reference-free correction is achieving robustness against the diverse failures and environmental factors seen in real-world deployments.
Thus, it is essential to use a dataset that contains numerous instances of systematic drift, failures, and other erratic behaviors known to hinder LCS correction when developing and validating correction models \cite{lcs_outdoor_survey}.

\begin{figure}[h]
    \centering
    \includegraphics[width=0.9\columnwidth]{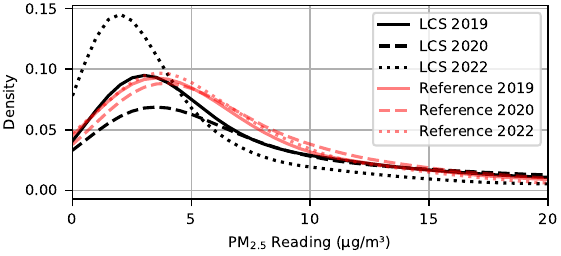} 
    \caption{Probability Density Function (PDF) of PM$_{2.5}$ readings from an LCS device co-located next to a reference station over 3 years. The PDF of the LCS readings matches the reference in the first year of deployment, then shows significant drift over the next two years, unlike the well-maintained reference station that exhibits consistent behavior.}
  \label{fig:distributionovertime}
\end{figure}

These combined real-world challenges often cause standard denoising and sensor fusion approaches, such as least-squares methods and Kalman filters \cite{kalman}, to fail as their state estimation becomes unreliable when readings are extremely erratic or contain missing values.
To further illustrate the issue of sensor drift, we show in Figure \ref{fig:distributionovertime} an example taken from AQ-SDR, shown as a comparison over a three-year period between a low-cost air quality sensor and a co-located, calibrated reference station in the city of Groningen in the Netherlands.
LCS exhibit a noticeable shift in the distribution of their PM$_{2.5}$ readings over the years, despite having a distribution similar to that of the reference station around its initial deployment in 2019.
In contrast, the well-maintained reference station shows consistent behavior, with a nearly identical data distribution over the same period.

To accurately model sensor bias, we build upon established findings that show LCS errors exhibit both nonlinear patterns and a systematic bias with heteroscedastic variance caused by environmental factors \cite{lcs_variance,lcs_outdoor_survey}. 
These error characteristics motivated the use of techniques like Gaussian process regression to correct LCS readings \cite{lcs_gaussian,variationalbias}.
Consequently, we adopt a similar rationale and propose a probabilistic sensor fusion model based on advances in Variational Autoencoders (VAEs) \cite{vae}.

\subsection{Model Overview}
\label{subsection:elimethod}

\begin{figure*}[!t]
  \centering
  \includegraphics[width=0.95\textwidth]{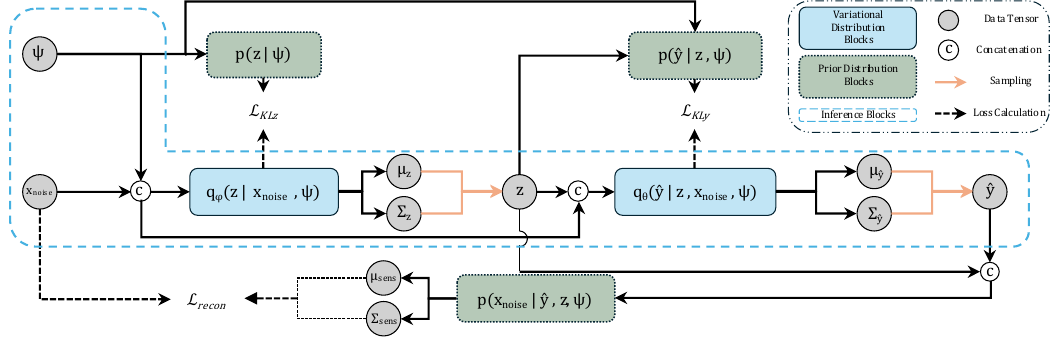}
  \caption{Veli structure following the derivation in Section \ref{subsection:lcsnoisemodel}. The input starts with AQ readings $x$ and auxiliary mask of `NA' readings $\psi$ on the left, propagating through the model's layers to generate a prediction of clean readings $\hat y$. 
  Conditioning on $\psi$ is omitted in some blocks for visual clarity but is implemented properly. 
  Prior distribution blocks (green) are used in the training to estimate the variational distribution blocks (blue), which are used in the inference as indicated by the blue dashed line. 
  All distribution blocks are modeled by two multilayer perceptron (MLP) layers followed by an MLP layer for each of the mean and variance. 
  The losses $\mathcal{L}_{KL_z}$, $\mathcal{L}_{KL_y}$, and $\mathcal{L}_{recon}$ correspond to the three terms in eq. \eqref{eq:final_loss}. Sampling refers to the traditional reparameterization in VAEs \cite{vae}.}
  \label{fig:modeldiagram}
\end{figure*}

Our probabilistic model, shown in Figure \ref{fig:modeldiagram}, is designed to separate the true AQ readings from sensor noise. It learns a mapping from a noisy high-dimensional input stream to a low-dimensional latent variable. This latent variable represents a fused reading on a continuous manifold, which facilitates the reconstruction of a clean, corrected output.
We enable the encoder to learn a robust mapping from any given reading to this manifold by training the model on a diverse range of noisy inputs.

In this implementation, we focus on correcting individual snapshots of LCS readings rather than modeling changes over time. We propose this design decision for two key reasons:
First, it is difficult to obtain perfectly time-aligned data streams from multiple adjacent sensors without encountering gaps or simultaneous failures. 
Second, simultaneously modeling time alongside all noise patterns (e.g., spikes, missing data) compromises the model's ability to capture diverse non-temporal noise patterns.
While our model processes hourly readings per pass, this snapshot-based approach does not discard the underlying temporal information. 
Since the correction model uses Lipschitz continuous layers (MLPs), temporal signatures in the corrected output remain preserved, as previously explored in \cite{lipschitz}. 
\subsection{LCS Noise Model}
\label{subsection:lcsnoisemodel}
In this section, we provide the necessary formulation to build Veli. We refer readers to \cite{vae} for more insights on the foundations of VAEs, and provide a more detailed derivation in Appendix \ref{appendix:derivation}.

To model the general structure of noisy LCS readings, we start by defining a basic distribution for LCS readings:
\begin{equation}
    {x}_{\rm noise}
    \;\sim\; 
    \mathcal{N}\!\Bigl(y + \mu_{\rm sens},\; \Sigma_{\rm sens}\Bigr)
    \label{eq:probabilistic_sensor_model}
\end{equation}
where ${x}_{\rm noise} \in \mathbb{R}^d$ is a noisy, raw AQ reading from $d$ different sensors in the same vicinity and $y \in \mathbb{R}^d$ is the unobserved AQ reading if it were measured by an ideal instrument (e.g., reference station). As stated earlier, we are building a reference-free method, so $y$ is inaccessible to us, and we replace it with predictions $\hat y$.
$\mu_{\rm sens} \in \mathbb{R}^d$ and positive definite diagonal covariance matrix $\Sigma_{\rm sens} = \operatorname{diag}(\sigma_{\rm sens,1}^2,\dots,\sigma_{\rm sens,d}^2)$ are non-constant, nonlinear bias and heteroscedastic terms that affect the LCS reading. 
While $\mu_{\rm sens}$ and $\Sigma_{\rm sens}$ do not model noise resulting from extreme spikes and missing data (extreme noise conditions), they can be used to produce a robust estimate of what the reading would be under normal noise conditions.

To enhance the representational capacity of the heteroscedastic terms, we introduce $z \in \mathbb{R}^r$ as a latent variable, where $r \leq d$. To model $z$, we condition it on an auxiliary parameter that contains additional information about the data, $\psi \in \mathbb{R}^d$. We then propose the following prior distribution: 
    \begin{align}
    \label{eq:pz_prior}
    p(z \mid \psi) = N(\mu(\psi),\Sigma(\psi))
    \end{align}
Standard VAEs typically use a standard Gaussian prior, $N(0,I)$.
However, to build a more identifiable and flexible prior, we follow the approach in \cite{ivae} and introduce $\psi$ as our auxiliary parameter.
This approach allows the latent space to effectively learn diverse variations within the input data, which is essential in filtering erratic behavior.
In the same manner, and since we are operating without LCS-reference-paired readings $(x_{\rm noise}, y)$, we treat $y$ as a latent variable whose prior distribution is given as:
\begin{align}
    \label{eq:py_prior}
    p(y \mid z, \psi) = N(\mu(z,\psi),\Sigma(z,\psi))
\end{align}

\subsection{Variational Approximations}
We aim to reconstruct the signal by separately generating the clean and noisy components of the reading.
We tackle this by maximizing a variational lower bound on $\log p(x_{\rm noise})$ that contains $z$ and $y$, conditioned on $\psi$, using the joint distribution factorization:
\begin{equation}
\nonumber
p(x_{\rm noise},y,z \mid \psi) = p(z \mid \psi)p(y \mid z, \psi)p(x_{\rm noise} \mid y,z, \psi)
\end{equation}
To estimate the distributions of $y$ and $z$ through the term $p(y,z \mid \psi)$, we will need to evaluate an intractable integral with no closed-form solution.
Therefore, we introduce approximate variational distributions similar to \cite{vae}, defined as:
\begin{align}
    \nonumber
    q_\phi(z \mid x_{\rm noise},\psi) &\approx p(z \mid x_{\rm noise},\psi) \\
    \nonumber
    q_\theta(y \mid z,x_{\rm noise},\psi) &\approx p(y \mid z,x_{\rm noise},\psi)
\end{align}
Under the Gaussian assumption, the posterior $q_\phi$ becomes:
\begin{equation}
    \begin{split}
    q_{\phi}\bigl(z \mid x_{\rm noise},\psi\bigr)=\mathcal{N}\!\bigl(\mu_{z}^{\phi},\Sigma_{z}^{\phi}\bigr)
    \end{split}
    \label{eq:encoder_dist}
\end{equation}
In practice, $\mu_{z}^{\phi}$ and $\Sigma_{z}^{\phi}$ are produced by an encoder network with two-branch outputs $f_{\phi},g_{\phi}$, respectively, such that $\mu_{z}^{\phi} \;=\; f_{\phi,\mu}(x_{\rm noise},\psi)$ and $\log \Sigma_{z}^{\phi} \;=\; g_{\phi}(x_{\rm noise},\psi)$.  
Similar to eq. \eqref{eq:encoder_dist}, we can define the parameterized posterior approximation $q_\theta$ as:
\begin{equation}
    q_\theta(y \mid z,x_{\rm noise},\psi)
    = \mathcal N\!\Bigl(
        \mu_{y}^\theta,
        \Sigma_{y}^\theta
        \Bigr)
    \label{eq:decoder_dist}
\end{equation}
and is parameterized by $\theta$ in the same manner such that $\mu_{y}^{\theta} \;=\; f_{\theta}(z,x_{\rm noise},\psi)$ and $\log \Sigma_{y}^{\theta} \;=\; g_{\theta}(z,x_{\rm noise},\psi)$. In this design, $\mu_{y}^\theta$ is the clean reading mean estimate $\hat y$.

Using eqs. \eqref{eq:decoder_dist} and \eqref{eq:encoder_dist}, we can approximate the intractable term $p(y,z \mid \psi)$ with a variational approximation $q_{\theta,\phi}(y,z \mid x_{\rm noise},\psi)$. Substituting $q_\phi$ and $q_\theta$ into the log-likelihood allows us to derive the Evidence Lower Bound (ELBO).
Minimizing the negative ELBO sets the objective to find optimal parameters $\phi, \theta$ for our model, such that:
\begin{align}
\nonumber
\log p(x_{\rm noise} \mid \psi)
\ge
\mathbb{E}&_{\,q_{\rm \theta,\phi}(y,z \mid x_{\rm noise},\psi)}\bigl[\log p(x_{\rm noise},y,z \mid \psi)
\\
\nonumber
& -\log q_{\rm \theta,\phi}(y,z \mid x_{\rm noise}, \psi)\bigr]
\end{align}
By using eq. \eqref{eq:probabilistic_sensor_model} as our reconstruction goal and incorporating $z$ and $\psi$ into the design, the final negative ELBO becomes:
\begin{align}
    \label{eq:final_loss}
    \mathcal{L}&(\theta, \phi)
    = \beta_{\rm z}D_{\rm KL}\bigl(q_\phi(z \mid x_{\rm noise}, \psi)\,\|\,p(z \mid \psi)\bigr)\\
    &\;+\;\nonumber
    \beta_{\rm y}D_{\rm KL}\bigl(q_\theta(y \mid z,x_{\rm noise}, \psi)\,\|\,p(y\mid z, \psi)\bigr)\\&
    +\nonumber
      \alpha
      \nonumber
      \sum_{i=1}^{d}
\left[
    \log\!\bigl(2\pi\,\sigma_{\mathrm{sens}}^2(z)_i\bigr)
    +\frac{\bigl(x_i - \hat y_i - \mu_{\mathrm{sens}}(z)_i\bigr)^2}
         {\sigma_{\mathrm{sens}}^2(z)_i}
\right]\nonumber
\end{align}
where $\sigma_{\mathrm{sens}}^2(z)_i$ and $\mu_{\mathrm{sens}}(z)_i$ are the non-linear bias and heteroscedastic terms in eq. \eqref{eq:probabilistic_sensor_model}, and $\hat y$ is the sampled prediction from the distribution in eq. \eqref{eq:decoder_dist} during training, but is taken as a point estimate during inference. 
Here $\alpha, \beta_{\rm z}, \beta_{\rm y} > 0$ are tunable coefficients similar to \cite{betavae}.

The goal of this formulation is to minimize the reconstruction term to shrink toward zero, which happens when 
$\mu_{\rm sens}(z)$ absorbs the noise that creates $x_{\rm noise}$, allowing $q_\theta$ to recover the underlying clean reading $y$. 
Concretely, $q_\theta(y \mid z, x_{\rm noise}, \psi)$ then learns the noise-free form of the signal such that our best estimate of the underlying clean signal, given a noisy input, becomes $\mu_{y}^{\theta}(z, x_{noise}, \psi)$.

This approach offers a clear advantage as it learns a smooth latent manifold that transforms erratic sensor readings, including missing values and spikes, into clean, continuous representations for decoding.
By training on a richly varied dataset, the model captures the full spectrum of AQ conditions, producing a latent encoding for virtually any combination of noisy readings.

\section{Experiments and Results}
\label{sec:experimentresult}
\subsection{AQ-SDR Dataset}
\label{subsection:AQ-SDR}
\subsubsection{Dataset Details}
\begin{figure}[h]
  \centering
  \includegraphics[width=0.9\columnwidth]{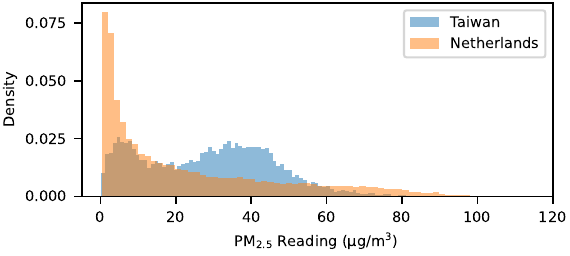}
  \caption{PDF Comparison of in-distribution and out-of-distribution data. Readings from the Netherlands are skewed to the left, indicating lower pollution levels, in contrast to the readings from Taiwan that reflect higher levels of pollution.}
  \label{fig:twnldists}
\end{figure}

To build our model, we use our proposed dataset, the AQ-SDR, which aggregates the LCS data from three major citizen-science initiatives: SamenMeten, Sensor.Community, and Location Aware Sensing System (LASS) community \cite{taiwan_data}.
To supply reference measurements to validate our method, we provide data from four authoritative sources: LuchtMeetNet (Air Measurement Network), the Royal Netherlands Meteorological Institute, the European Environment Agency, and the Taiwanese Ministry of Environment Open Data.
The majority of the deployed LCS began operating in 2019 and continue to provide measurements, with a smaller subset having been operational since before 2019.
To further evaluate our model's generalizability across different pollution levels and regions, we create two partitions of the AQ-SDR dataset: an in-distribution set and an out-of-distribution set.
We illustrate the difference between the in-distribution data and out-of-distribution data using two samples shown in Figure \ref{fig:twnldists}. 
The left-skewed distribution of the in-distribution data (Netherlands) reflects lower pollution levels, in contrast to the right-skewed distribution of the out-of-distribution data (Taiwan), which indicates higher pollution levels.

The first, in-distribution partition of the AQ-SDR has data from 99 sites across the Netherlands, each location hosting ten different LCS with no co-located reference station (reference-free). Its corresponding test set has five sites, chosen to provide the widest geographic coverage possible, each co-located with at least one reference station.
The out-of-distribution partition consists of data from 55 heavily polluted locations in Taiwan, with no co-located reference station. The test set consists of five locations with co-located reference stations, similarly chosen to provide a wide geographic coverage.
The dataset and the code to generate the partitions will be made publicly available, and we report further details on the dataset in Appendix \ref{appendix:dataset}. AQ-SDR will also be made accessible to the public through an interactive online dashboard as shown in Figure \ref{fig:hourly-avgs-utrecht}.

\subsubsection{Dataset Processing}
Each LCS site hosts ten temporally aligned PM$_{2.5}$ sensors, which are sampled hourly, such that:
\[
\bigl\{\,x_{i}(t)\;\bigm|\;i=1,\dots,10,\;t=t_{1},\dots,t_{T}\bigr\},
\]
where $x_{i}(t)$ is the reading from the $i$-th sensor at time $t$. While using ten sensors was our design choice, we show in Section \ref{subsection:viablesensors} implementations with fewer than ten sensors.
To model missing data (referred to as `NA' in this work) for sensor $i$ at time $t$, we define an auxiliary mask $\psi$ that is aligned in time to each location, such that $\psi_{i}(t) = 1$ if the data is observed, and $\psi_{i}(t) = 0$ if the data is missing (`NA').
This mask is essential for modeling data absence as conditional information in our model.
In evaluation regions, we have aligned reference (ground truth) data $y(t)$.
If there is more than one nearby reference station, we average their readings. 
At no point during the training was the model exposed to reference readings, keeping it completely reference-free.

\begin{figure*}[!t]
  \centering
  \includegraphics[width=\textwidth]{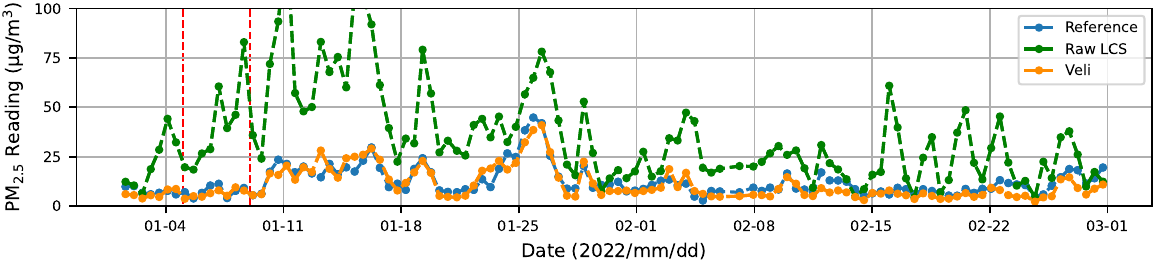}
  \caption{12-hour-averages for Utrecht's data over two months. The readings of the raw LCS deviate significantly from the reference reading. Veli takes these readings as an input and outputs an accurate corrected measurement that matches the reference's readings. The region in the red-dashed lines is zoomed in on Figure \ref{fig:hourly-avgs-utrecht}.}
  \label{fig:daily-avgs-utrecht}
\end{figure*}

\subsection{Implementation Details \& Evaluation Metrics}
\label{subsection:implementationdetails}

The model was implemented using PyTorch 2.3.1 and trained on an NVIDIA RTX 3090 GPU. We trained the model for 100 epochs with an ADAM optimizer, a batch size of 64, and an initial learning rate of $1\times10^{-6}$.
Hyperparameters $\alpha,\beta_{\rm z}, \beta_{\rm y}$ in eq. \eqref{eq:final_loss} are set to 1, 10, 0.1, respectively, and we provide sensitivity analysis in Appendix \ref{subsection:lossweights}.



All MLP layers that do not concatenate inputs use a hidden dimension of 32.
For out-of-distribution fine-tuning, we froze the decoder and trained only the encoder for an additional 30 epochs on the new data distribution. 
Other implementation, tensor preparation, and evaluation details match previous works in time-series modeling \cite{liu2024itransformer}.

To evaluate our model, we use Mean Absolute Error (MAE) as the standard metric from the literature to compare model outputs to reference readings \cite{Concas2021LowCost,survey2}. We also report the average and standard deviation of the output over five runs with different random seeds for all numerical results. 
In all experiments, `raw LCS' is the input to Veli.

As this is the first work to propose a completely reference-free correction method, no other methods exist for a direct comparison. In addition, no large-scale unifying benchmark exists aside from AQ-SDR. We instead provide a comparison against traditional blind denoising techniques like Kalman Filters (KF) \cite{kalman} and Principal Component Analysis (PCA) denoising \cite{pca}. A KNN imputer was used to enable these two methods to run on data with missing readings. In addition, we provide extended results and analysis in Appendix \ref{appendix:ablation}.

\subsection{Correction Results}
\label{subsection:correctionresults}
\subsubsection{In-distribution Results}
Table \ref{tab:errors_indist} presents the model's performance across five locations in five different cities in the Netherlands. 
The MAE decreased substantially compared to the raw LCS readings in Amsterdam, Rotterdam, and Utrecht.
We also show the LCS units from IJmuiden and Nijmegen providing accurate readings that do not require correction. Veli introduces minimal stochastic noise due to sampling, and we expand on this in Appendix \ref{subsection:underestimation}.

\begin{table}[h]
  \centering
  \begin{tabular}{ccccc}
    \toprule  
    \multicolumn{1}{c|}{\multirow{2}{*}{City}} & \multicolumn{4}{c}{MAE ($\mu\mathrm{g}/\mathrm{m}^3$)} \\
    \cmidrule(lr){2-5} 
    
    
    \multicolumn{1}{c|}{} & \multicolumn{1}{c}{LCS$_{m}$} & \multicolumn{1}{c}{PCA} & \multicolumn{1}{c|}{KF} &\multicolumn{1}{c}{Veli}   \\
    \midrule
    
    \multicolumn{1}{l|}{Amsterdam}  & 11.34 & 10.45 & \multicolumn{1}{c|}{9.77}   & 3.73$\pm$0.15      \\
    \multicolumn{1}{l|}{Rotterdam}  & 21.27 &22.31 & \multicolumn{1}{c|}{11.57}   & 3.36$\pm$0.37    \\
    \multicolumn{1}{l|}{Utrecht}    & 24.77 & 13.72 & \multicolumn{1}{c|}{15.95}   & 5.25$\pm$0.26     \\
    \multicolumn{1}{l|}{IJmuiden}   &  4.02  & 3.93 &\multicolumn{1}{c|}{4.36}   &  3.44$\pm$0.20      \\
    \multicolumn{1}{l|}{Nijmegen}   &  2.82  & 2.82 & \multicolumn{1}{c|}{2.96}   &  3.06$\pm$0.18     \\
    \bottomrule
  \end{tabular}
  \caption{MAE comparison for in-distribution raw LCS, PCA denoising, KF denoising, and Veli's output. LCS$_{m}$ is the average of raw LCS readings. Veli's results show mean $\pm$ standard deviation across five random seeds.}
  \label{tab:errors_indist}
\end{table}

Figure \ref{fig:daily-avgs-utrecht} presents a 12-hour-average time series over a two-month period in Utrecht, which has the worst raw LCS accuracy among the selected regions.
The region in the graph between the red-dashed lines highlights a four-day window, whose hourly sampling was shown earlier in Figure \ref{fig:hourly-avgs-utrecht}.
Veli successfully captures both short- and long-term trends and spikes, despite our model being completely reference-free.
While our model performs best when the raw LCS data exhibits an underlying trend (i.e., is not completely random), it is also designed to handle a common failure mode with random readings, demonstrated in Section \ref{subsection:ablationstudy}.

\begin{figure}[h]
    \centering
    \includegraphics[width=0.9\columnwidth]{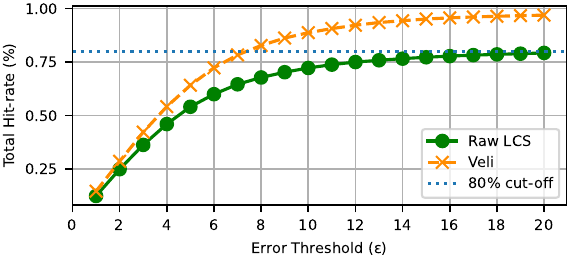} 
    \caption{Percentage of data points that are within a threshold $\epsilon$ of the reference readings.}
  \label{fig:hitrateutrecht}
\end{figure}
To obtain a holistic view of Veli's correction, we measure the hit rate, defined as the percentage of individual readings whose value is within a given threshold ($\epsilon$) from the reference value. We count all measurements that have MAE $\leq \epsilon$, and plot this percentage of total data in Figure \ref{fig:hitrateutrecht}.
For the raw LCS readings, we need to relax the MAE margin to be up to 20 to capture 80\% of all readings, in contrast to our model, for which the required margin is reduced to only 7.34.

\subsubsection{Out-of-distribution Results}
We further evaluate our model on out-of-distribution data from five locations in five different cities in Taiwan, shown in Table \ref{tab:erroroutofdist}.
``Veli zero-shot'' denotes applying the weights trained on in-distribution data directly. For the fine-tuning variant, we froze the decoder and trained the encoder for 30 additional epochs on the Taiwanese LCS subset (reference-free). 
While the model shows strong average performance in a zero-shot setting, the results are inconsistent, illustrated by the high standard deviation across experiments. After fine-tuning, the model becomes significantly more reliable on out-of-distribution data.


\begin{table}[h]
  \centering
  \setlength{\tabcolsep}{3pt} 
  \begin{tabular}{c| c c c| c| c}
    \toprule
    \multirow{2}{*}{City} & \multirow{2}{*}{LCS$_{m}$} & \multirow{2}{*}{PCA} & \multirow{2}{*}{KF} & \multicolumn{2}{c}{Veli} \\
    \cline{5-6}
     & & & & Zero-shot & Fine-tuned \\
    \midrule
    Taichung & 10.01 & 10.01 & 9.98  & 7.78$\pm$1.22 & 7.65$\pm$0.03 \\
    Tainan   & 14.09 & 14.25 & 13.28 & 8.59$\pm$1.48 & 7.83$\pm$0.27 \\
    Taoyuan  & 9.22  & 9.11  & 9.04  & 5.79$\pm$0.10 & 5.64$\pm$0.06 \\
    Taipei   & 7.52  & 7.49  & 7.58  & 6.51$\pm$0.98 & 6.43$\pm$0.03 \\
    Puzi     & 13.75 & 13.70 & 13.80 & 9.10$\pm$1.27 & 9.04$\pm$0.09 \\
    \bottomrule
  \end{tabular}
  \caption{MAE ($\mu\mathrm{g}/\mathrm{m}^3$) comparison for out-of-distribution raw LCS, PCA denoising, KF denoising, and Veli's output. LCS$_{m}$ is the average of raw LCS readings. Veli's results show mean $\pm$ standard deviation across five random seeds.}
  \label{tab:erroroutofdist}
\end{table}

\subsection{Model Analysis and Discussion}

\subsubsection{Temporal Analysis}
\begin{figure}[h]
    \centering
    \includegraphics[width=0.9\columnwidth]{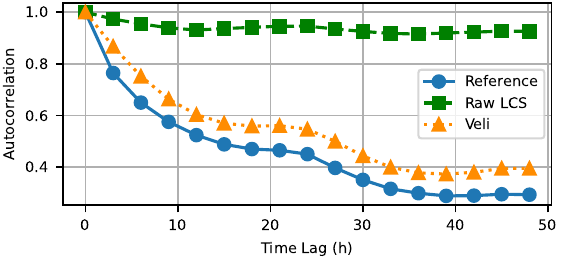} 
    \caption{Comparison of autocorrelation over 48 hours. Correcting the raw LCS with Veli yields a behavior that is similar to a well-maintained reference station.}
  \label{fig:autocorrelation}
\end{figure}

PM$_{2.5}$ readings typically exhibit strong autocorrelation that gradually decays over time, primarily driven by the underlying pollutant concentrations \cite{zaini_pm25_2022}.
In addition to this inherent structure, noise from LCS can also introduce significant autocorrelation, often persisting for up to 48 hours, as illustrated in Figure \ref{fig:autocorrelation}.
The raw LCS readings remain highly autocorrelated for an extended period of time, in contrast to the trends seen in the reference stations.
Our model eliminates this noise, producing outputs that closely match the reference time series.
This behavior is consistent with the discussion in Section \ref{subsection:problemformulation}, showing that our model corrects the readings without compromising temporal information.
\subsubsection{Ablation Studies \& Limitations}
\label{subsection:ablationstudy}
\begin{figure}[h]
    \centering
    \includegraphics[width=0.9\columnwidth]{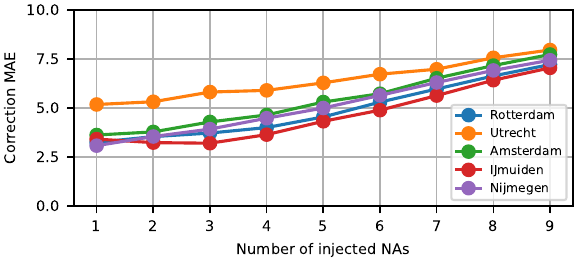} 
    \caption{Effect of modeling sensor failure by injecting `NA' readings into available LCS readings.}
  \label{fig:nainjection10}
\end{figure}

To simulate adversarial sensor failure (channel dropout), we took the original test data and randomly replaced a fixed number of the 10 sensor readings, $n$, with `NA' values for each hourly sample.
We then evaluated the model's performance for different values of $n$, from 1 to 9.
Figure \ref{fig:nainjection10} shows how these injected failures degrade the correction's performance, but remains within an acceptable range of accuracy (MAE $<$ 10). For sensors that are already accurate (e.g., Nijmegen's LCS), using a variation of Veli with fewer channels would be beneficial, which we show in the next subsection and further in Appendix \ref{subsection:syntheticdata}.



\subsubsection{Minimum Number of Viable Sensors}
\label{subsection:viablesensors}

\begin{figure}[h]
    \centering
    \includegraphics[width=0.95\columnwidth]{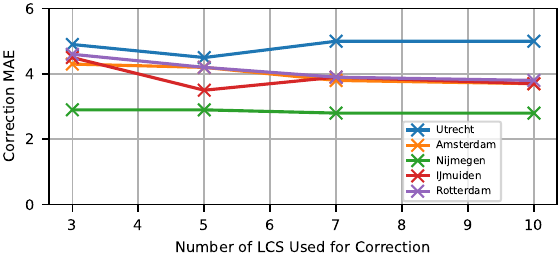} 
    \caption{MAE of applying Veli on LCS readings when trained and tested on subsets with 3, 5, 7, and the default 10 sensors. The results demonstrate that performance is not significantly impacted by a reduction in sensor count.}
  \label{fig:viablesensors}
\end{figure}

As established previously, our configuration uses a collection of 10 LCS per region.
To evaluate Veli's flexibility, we tested its correction performance on subsets containing only 3, 5, and 7 sensors.
For every sample, we ensured that at least half the sensors had a non-NA reading (rounded down).
As Figure \ref{fig:viablesensors} shows, reducing the number of sensors does not significantly affect Veli's performance.
However, using only three sensors increases the risk of connectivity loss, which can result in data gaps. Therefore, we retain 10 sensors as our standard configuration to maximize connectivity and data availability, but show that Veli remains effective using as few as 3 sensors.

\section{Conclusion}
\label{sec:conclusion}
In this work, we presented Veli, an unsupervised Bayesian correction method for low-cost AQ sensors that does not require high-cost reference stations, lowering the barrier for deploying dense monitoring networks.
To develop and evaluate the model and to build a unifying benchmark for AQ monitoring, we also presented AQ-SDR, the largest AQ benchmark to date.
AQ-SDR contains data from 23,737 sensors distributed across multiple regions, capturing a diverse set of sensor errors and failure modes.
Our comprehensive evaluation demonstrates that Veli provides robust correction across varying pollution levels and data distributions. Our proposed model is resilient against common failures, such as erratic spikes and complete sensor blackouts.
We envision this work serving both as a practical solution for long-term LCS deployment and as a foundational benchmark to pave the way for future research in AQ monitoring.

\section{Acknowledgements}
We sincerely thank the Dutch government for supporting this research with the starter grant (startersbeurzen). We also thank the organizations and researchers who provide the open data to enable this research, including the Dutch National Institute for Public Health and the Environment (RIVM), the Dutch Royal Netherlands Meteorological Institute (KNMI), Dr. Ling-Jyh Chen in Taiwan Academia Sinica for the AirBox project, the Taiwan Ministry of Environment, the Sensor.Community platform, and the European Environmental Agency (EEA). We also thank the GGD Amsterdam and RIVM for providing information about how air quality sensor stations work in the Netherlands. We also thank the CREATE Lab at the Robotics Institute at Carnegie Mellon University for the technical support in building the air quality dashboard.

\appendix
\section{ELBO Derivation}
\label{appendix:derivation}

We assume the following factorization of the joint and the conditional distributions:
\begin{equation}
\label{eq:factorization}
\begin{aligned}
&p(x,y,z)
\;=\;
p(z)\
p(y \mid z)\
p(x \mid y,z)\\
&p(y,z \mid x)
\;=\;
p(z \mid x)\
p(y \mid z,x)
\end{aligned}
\end{equation}
To simplify the notation, we use $x$ to refer to $x_{\rm noise}$ in the main text. We show all derivations for $y,x \in \mathbb{R}^d$ and $z \in \mathbb{R}$, but they can be extended to any vector $z \in \mathbb{R}^r$ where $r \leq d$. We also show the derivation without conditioning on $\psi$, but it remains valid in both cases. 
We reserve $q$ to denote variational approximations for probability distributions $p$. In particular, we propose a variational approximation for $p(y,z \mid x)$ parameterized by a neural network with parameters $(\theta, \phi)$ such that
\[
q_\phi(z \mid x)\approx p(z \mid x),
\quad
q_\theta(y \mid z,x)\approx p(y \mid z,x)
\]
\[
q_{\rm \theta,\phi}(y,z \mid x) = q_\phi(z \mid x) q_\theta(y \mid z,x) 
\]

\bigskip

\subsection{Standard ELBO Identity}

We rewrite $p(x,y,z)$ as:
\[
p(x)
\;=\;
\frac{p(x,y,z)}{p(y,z \mid x)}
\]
Following the standard VAE formulation \cite{vae}, we take the logarithm and introduce the variational posterior to define the ELBO. The objective is to maximize this bound to obtain a tractable lower bound on the log-likelihood of the data, such that:

\begin{flalign*}
&\log p(x)
\;\ge\;
\mathbb{E}_{\,q_{\rm \theta,\phi}(y,z \mid x)}\bigl[\log p(x,y,z)
-\log q_{\rm \theta,\phi}(y,z \mid x)\bigr]
&
\end{flalign*}
\begin{flalign*}
\operatorname{ELBO}(\theta, \phi)
&=\mathbb{E}_{q_{\theta,\phi}(y,z \mid x)}\bigl[\log p(x,y,z)\bigr] \\
&- \mathbb{E}_{q_{\theta,\phi}(y,z \mid x)}\bigl[\log q_{\theta,\phi}(y,z \mid x)\bigr]
\end{flalign*}
Substituting the factorized terms in eq. \eqref{eq:factorization} and their posterior approximations, we obtain:
\begin{align*}
&\operatorname{ELBO}(\theta, \phi)
=\\
&\quad\; \mathbb{E}_{q_\phi(z \mid x)q_\theta(y \mid z,x)}
\Bigl[
\log\bigl(p(z)p(y \mid z)p(x \mid y,z)\bigr)
\Bigr]
\nonumber\\[-0.5ex]
&
-\mathbb{E}_{q_\phi(z \mid x)q_\theta(y \mid z,x)}
\Bigl[
\log\bigl(q_\phi(z \mid x)q_\theta(y \mid z,x)\bigr)
\Bigr]
\nonumber\\
&=
\underbrace{\mathbb{E}_{q_{\rm \theta,\phi}(y,z \mid x)}\bigl[\log p(x \mid y,z)\bigr]}_{\text{(A)}} \\
&+
\underbrace{\mathbb{E}_{q_{\rm \theta,\phi}(y,z \mid x)}\bigl[\log p(y \mid z)\bigr]}_{\text{(B)}}
+
\underbrace{\mathbb{E}_{q_\phi(z \mid x)}\bigl[\log p(z)\bigr]}_{\text{(C)}}
\nonumber\\
&
-\underbrace{\mathbb{E}_{q_{\rm \theta,\phi}(y,z \mid x)}\bigl[\log q_\theta(y \mid z,x)\bigr]}_{\text{(D)}}
-\underbrace{\mathbb{E}_{q_\phi(z \mid x)}\bigl[\log q_\phi(z \mid x)\bigr]}_{\text{(E)}}
\end{align*}
\subsection{KL penalties}
By grouping terms (B) and (D), we obtain:
\begin{align*}
&\mathbb{E}_{q_{\theta,\phi}(y,z|x)}[\log p(y|z)] - \mathbb{E}_{q_{\theta,\phi}(y,z|x)}[\log q_\theta(y|z,x)]\\
= & \mathbb{E}_{q_\phi(z|x)}\left[\mathbb{E}_{q_\theta(y|z,x)}[\log p(y|z) - \log q_\theta(y|z,x)]\right]\\
= -&\mathbb{E}_{q_\phi(z|x)}\left[D_{\rm KL}(q_\theta(y|z,x) \| p(y|z))\right]
\end{align*}

and grouping (C) and (E) gives us:
\begin{align*}
&\mathbb{E}_{\,q_\phi(z \mid x)}\bigl[\log p(z)\bigr]
-\,\mathbb{E}_{\,q_\phi(z \mid x)}\bigl[\log q_\phi(z \mid x)\bigr]
\nonumber\\[-0.5ex]
&\quad
=
-\,D_{\rm KL}\bigl(q_\phi(z \mid x)\,\|\,p(z)\bigr).
\end{align*}
In our implementation, we sample latent variables $z$ and $y$ using the reparameterization trick similar to VAEs. We also sample multiple values of $(z,y)$ such that the expectations $\mathbb{E}_{\rm q_{\rm \theta,\phi}(y,z \mid x)}$ and $\mathbb{E}_{\rm q_\phi(z \mid x)}$ are estimated per batch, allowing us to omit the expectation in the final ELBO expression. Combining these terms, we obtain the final ELBO term:

\begin{flalign*}
\operatorname{ELBO}(\theta, \phi) &= \mathbb{E}_{\,q_{\rm \theta,\phi}(y,z \mid x)}\Bigl[ \log p(x \mid y,z) \Bigr] \\
& - D_{\rm KL}\bigl(q_\phi(z \mid x)\,\|\,p(z)\bigr)\\
& - D_{\rm KL}\bigl(q_\theta(y \mid z,x)\,\|\,p(y \mid z)\bigr)
\end{flalign*}

Here, the term $\log p(x \mid y,z)$
rewards how well \(q_\theta(y,z\mid x)\) explains the data \(x\) via \(p(x\mid y,z)\), the term $D_{\rm KL}\bigl(q_\phi(z \mid x)\,\|\,p(z)\bigr)$ penalizes deviations of \(q_\phi(z\mid x)\) from the assumed prior \(p(z)\), and the term $D_{\rm KL}\bigl(q_\theta(y \mid z,x)\,\|\,p(y \mid z)\bigr)$ penalizes deviations of \(q_\theta(y \mid z,x)\) from \(p(y\mid z)\). All priors are assumed to be from Gaussian families.

\bigskip

\subsection{ELBO with Reconstruction Term}
Based on the heteroscedastic form we adopted for $p(x \mid y,z)$ in eq. \eqref{eq:probabilistic_sensor_model}, we substitute the values for mean and variance into the Gaussian log-probability density function to obtain:

\begin{align*}
      &\log p\bigl(x \mid y, z\bigr)
      \;=\;\\
      &-\,\frac{1}{2}
      \sum_{i=1}^{d}
      \Bigl[
        \frac{\bigl(x_i - (\hat y_i + \mu_{\mathrm{sens}}(z)_i)\bigr)^2}{\sigma^2_{\mathrm{sens}}(z)_i}
        +\log\bigl(2\pi\,\sigma^2_{\mathrm{sens}}(z)_i\bigr)
      \Bigr]
\end{align*}

Following \cite{betavae}, we introduced three hyperparameters $\alpha, \beta_{\rm z}$, and $\beta_{\rm y}$ to weight each term. The final minimization target, the negative ELBO, becomes:

\begin{align*}
\mathcal{L}&(\theta, \phi)
=\\
  &\alpha
  \sum_{i=1}^{d}
      \Bigl[
        \frac{\bigl(x_i - (\hat y_i + \mu_{\mathrm{sens}}(z)_i)\bigr)^2}{\sigma^2_{\mathrm{sens}}(z)_i}
        +\log\bigl(2\pi\,\sigma^2_{\mathrm{sens}}(z)_i\bigr)
      \Bigr]\\
&\;+\;
\beta_{\rm z}D_{\rm KL}\bigl(q_\phi(z \mid x)\,\|\,p(z)\bigr)\\
&\;+\;
\beta_{\rm y}D_{\rm KL}\bigl(q_\theta(y \mid z,x)\,\|\,p(y\mid z)\bigr)
\end{align*}

We also note that the same term holds by conditioning on $\psi$ (i.e., the joint distribution is $p(x,y,z \mid \psi)$) where $\psi$ is an additional variable that contains environmental information. This yields the loss term in eq. \eqref{eq:final_loss}, written here as:
\begin{align*}
\mathcal{L}&(\theta, \phi)
=\\
  &\alpha
  \sum_{i=1}^{d}
      \Bigl[
        \frac{\bigl(x_i - (\hat y_i + \mu_{\mathrm{sens}}(z)_i)\bigr)^2}{\sigma^2_{\mathrm{sens}}(z)_i}
        +\log\bigl(2\pi\,\sigma^2_{\mathrm{sens}}(z)_i\bigr)
      \Bigr]\\
&\;+\;
    \beta_{\rm z}D_{\rm KL}\bigl(q_\phi(z \mid x, \psi)\,\|\,p(z \mid \psi)\bigr)\\
    &\;+\;
    \beta_{\rm y}D_{\rm KL}\bigl(q_\theta(y \mid z,x, \psi)\,\|\,p(y\mid z, \psi)\bigr)
\end{align*}

\section{Additional Experiments \& Ablation}
\label{appendix:ablation}




\subsection{Performance on Synthetic Data}
\label{subsection:syntheticdata}

To stress-test our unsupervised method, we generate synthetic data streams and apply our model to them. Each synthetic stream is created from a ground-truth base signal, which is one of the following:
\begin{enumerate}
    \item A clean, one-year reference station reading (from Utrecht).
    \item A sinusoidal signal (offset = 2, maximum value = 30, period = 48 hours).
    \item A sawtooth signal (offset = 2, maximum value = 30, period = 48 hours).
    \item Randomly generated data from an exponential distribution with $\lambda = \frac{1}{12}$.
\end{enumerate}
We then apply noise to the base signal using a combination of four types:
\begin{itemize}
    \item Gaussian noise: A sample from $N(\mu,\sigma^2)$ is added to the base signal.
    \item Multiplicative noise: The base signal is multiplied by a small factor.
    \item Spike noise: The base signal is multiplied by a large spike factor.
    \item NA Reading: A reading is replaced with `NA'.
\end{itemize}
\begin{figure}[h!]
    \centering
    \includegraphics[width=0.95\columnwidth]{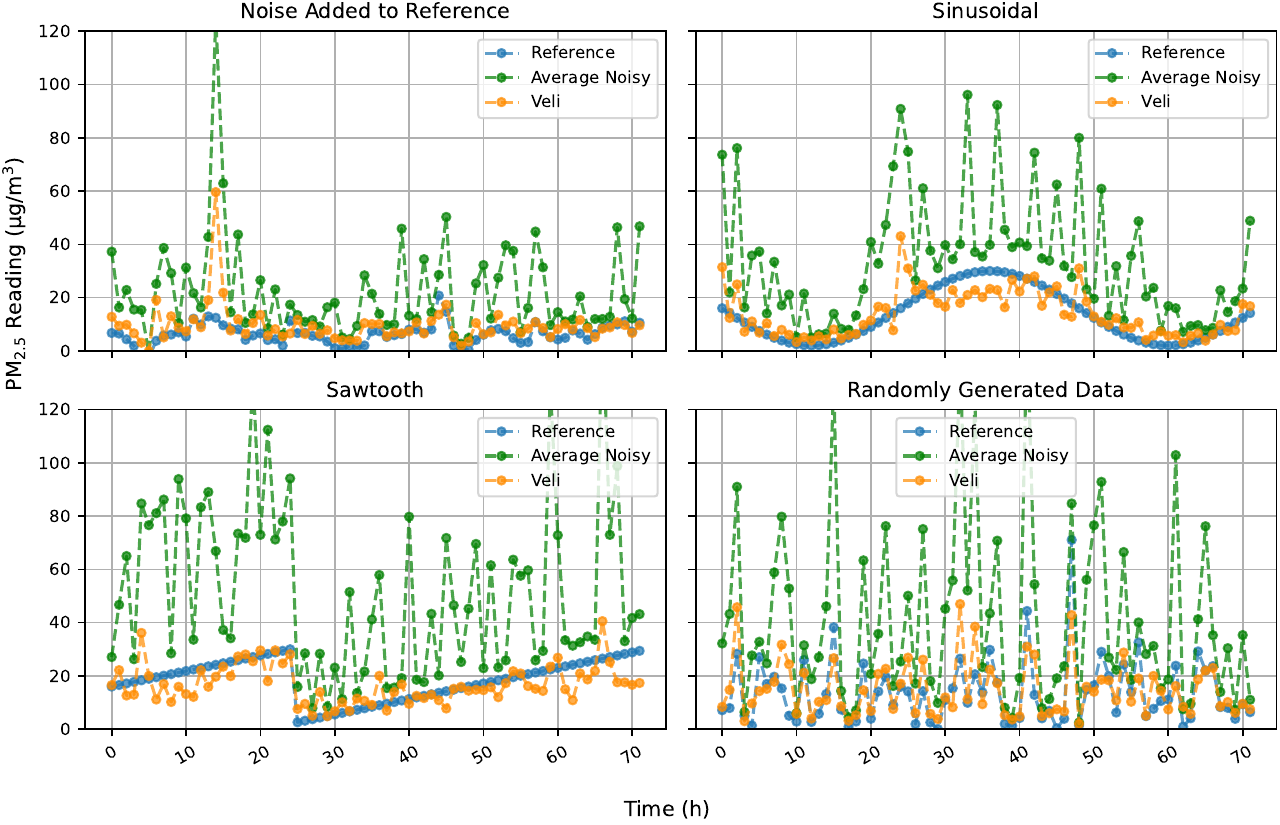} 
    \caption{Correction results on synthetic data with similar noise to AQ-SDR. Veli demonstrates robust correction despite working with different base signals. Noise parameters: Gaussian noise, $\mu = 3, \sigma = 2$, $p_{\text{Gaussian}}=1.0$ present. Multiplicative noise factor is 1.5, $p_{\text{factor}} = 0.5$. Spike factor is 10.0, $p_{\text{spike}} = 0.1$. No more than 5 `NA' per row are dropped with $p_{\text{NA}} = 0.35$.}
  \label{fig:syntheticsoft}
\end{figure}
\begin{figure}[h!]
    \centering
    \includegraphics[width=0.95\columnwidth]{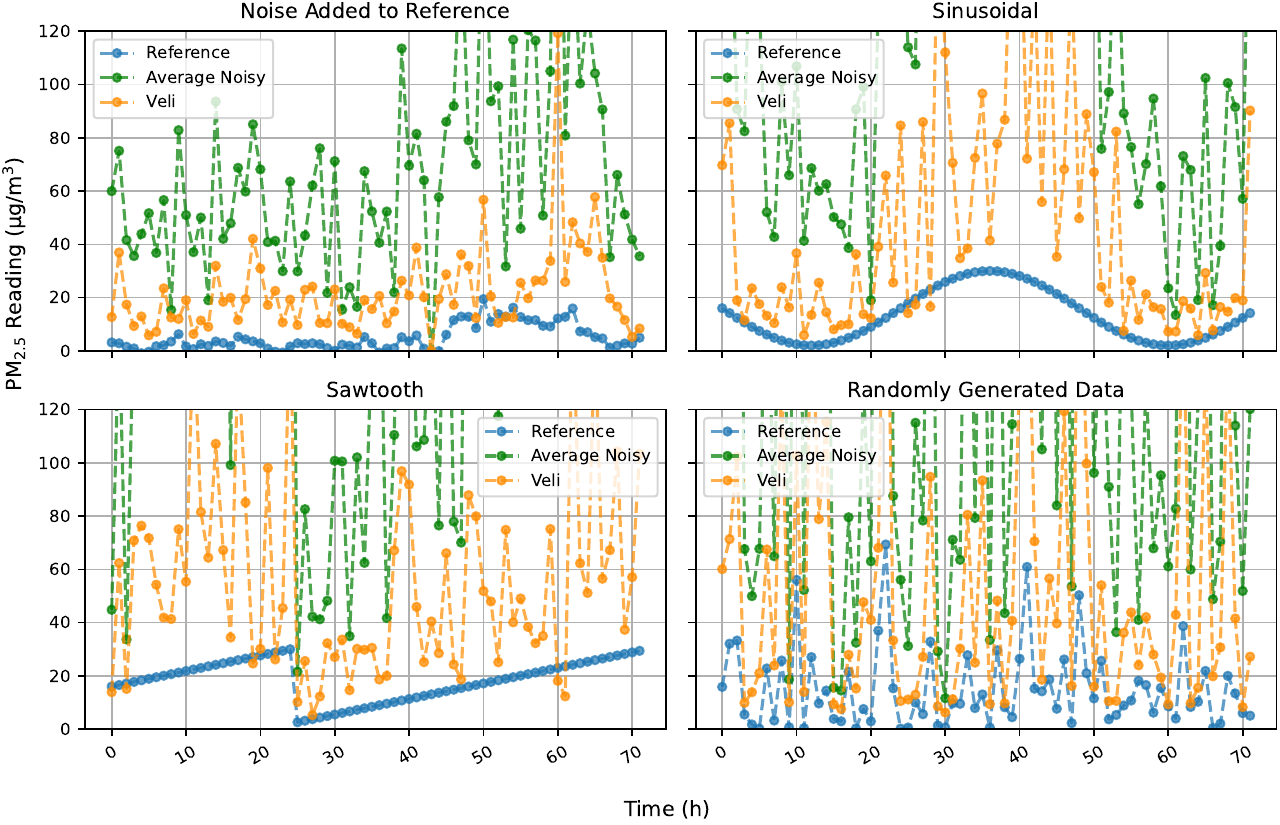} 
    \caption{Correction results on synthetic data with extreme noise. This shows a failure case where the data is so distorted that Veli cannot restore the base signal. Noise parameters: Gaussian noise, $\mu = 5, \sigma = 2$, $p_{\text{Gaussian}}=1.0$ present. Multiplicative noise factor is 2.0, $p_{\text{factor}} = 0.7$. Spike factor is 10.0, $p_{\text{spike}} = 0.4$. No more than 5 `NA' per row are dropped with $p_{\text{NA}} = 0.4$.}
  
  \label{fig:syntheticharsh}
\end{figure}
These noise sources are applied with probabilities $p_{\text{Gaussian}}$, $p_{\text{factor}}$, $p_{\text{spike}}$, and $p_{\text{NA}}$, respectively.
These noise sources are applied to each data point in combination according to their respective probabilities (probabilities are independent of each other). This noise generation process was identical for all ten synthetic streams.

Figures \ref{fig:syntheticsoft} and \ref{fig:syntheticharsh} show two key results from this simulation. The first figure demonstrates a noise profile designed to mimic the noise behavior of the LCS in the AQ-SDR dataset. The results show our model performs robustly, even when the underlying base signals do not resemble typical AQ data.
We note that while our model captures the base signal, residual noise characteristics can still be observed in the final corrected output.
The second figure shows a scenario with harsher noise, where the data is distorted beyond recognition. 
This scenario demonstrates a clear failure case, as the model cannot recover the original signal.

\subsection{Predictions with Credible Intervals}
\label{subsection:credibleinterval}
\begin{figure}[h]
    \centering
    \includegraphics[width=0.9\columnwidth]{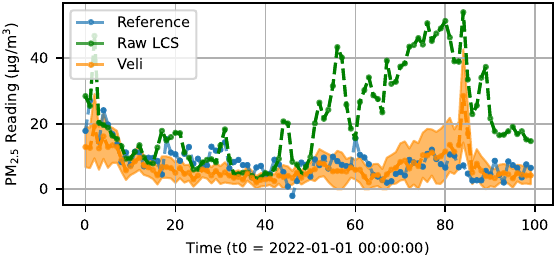} 
    \caption{Time series for Utrecht's data over four days. The shaded orange region is a one-standard-deviation credible interval generated from the model's output ($\hat y \pm \sqrt{(\Sigma_y^\theta)_{ii}}$).}
  \label{fig:timecredible}
\end{figure}

In this section, we investigate our model's ability to generate a credible interval from our Bayesian framework.
As discussed in Section \ref{subsection:elimethod}, our Bayesian model outputs predictions by sampling from a Gaussian distribution. In all analyses done in inference, the output was the point estimate $\hat y = \mu_y^\theta(z, x_{noise}, \psi)$, discarding the effect of the diagonal covariance matrix, $\Sigma_y^\theta$. The standard deviation for every output channel would then be $\sqrt{(\Sigma_y^\theta)_{ii}}$. 
Figure \ref{fig:timecredible} shows the effect of implementing a credible interval within one standard deviation of the prediction. 
We contrast this figure with using only a point estimate, as shown earlier in a different time period in Figure \ref{fig:hourly-avgs-utrecht}.
We observed that extreme deviations (spikes) or numerous missing values (`NA') resulted in higher standard deviations, which is our framework's response to uncertainty.
Although credible intervals yield improved results, they are not a standard method for model evaluation.
Therefore, we present this capability as an additional benefit of our framework rather than including it in the main analysis. For example, a large deviation for an extended period of time could be used to indicate sensor failure or abnormal conditions.

\subsubsection{Limitations of Reference-Free Learning}
\label{subsection:underestimation}

\begin{figure}[h]
    \centering
    \includegraphics[width=0.9\columnwidth]{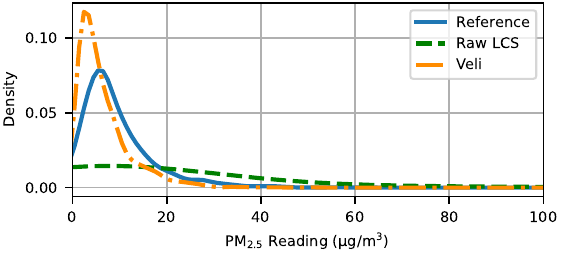} 
    \caption{PDF of the aggregate of readings from each source. The raw LCS shows one of the most erratic sensors in AQ-SDR. Veli's output distribution from these sensors closely matches the reference readings, but still underestimates their true values, demonstrated by its left-skewed peak.}
  \label{fig:distributioncorrection}
\end{figure}

A noticeable limitation of our proposed method is underestimation of the true reference values. Figure \ref{fig:distributioncorrection} shows the distributions of raw LCS readings, Veli's outputs, and reference values. 
Although our model successfully closely approximates the underlying readings, there is a pattern of underestimation when compared to the reference readings. This underestimation of approximately 3 $\mu\mathrm{g}/\mathrm{m}^3$ is observed only for air quality readings below 15 $\mu\mathrm{g}/\mathrm{m}^3$, an error that is not significant for hazardous event prediction. We also point out that our proposed credible intervals provide a robust mitigation strategy against this limitation.
This underestimation is a consequence of the unsupervised approach. Unlike previous reference-based works that use reference readings to correct LCS readings \cite{maximizingcorrelation}, Veli does not have a ground-truth signal to guide the output distribution toward the true data distribution.

\subsection{Loss Weights}
\label{subsection:lossweights}

\begin{figure}[h]
    \centering
    \includegraphics[width=0.95\columnwidth]{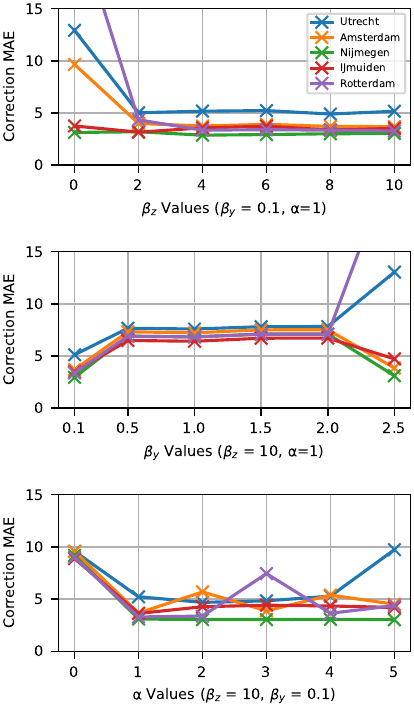} 
    \caption{Result of varying the hyperparameters $\alpha$, $\beta_z$, and $\beta_y$. Default values are 1, 10, 0.1, respectively.}
  \label{fig:lossweights}
\end{figure}

Our loss function, defined in eq. \eqref{eq:final_loss}, is controlled by three hyperparameters: $\alpha$, $\beta_z$, and $\beta_y$. 
The hyperparameters $\beta_z$ and $\beta_y$ regulate the influence of the priors, while $\alpha$ weights the reconstruction of the noisy signal. 
We present in Figure \ref{fig:lossweights} an ablation that empirically explores the sensitivity of our model to hyperparameter variations.

The hyperparameters are scaled relative to their final values ($\alpha = $1, $\beta_z = $ 10, and $\beta_y =$ 0.1), with the exception of $\beta_y$, which is kept above zero. Setting $\beta_y = 0$ causes the loss to diverge, which destabilizes the training process.
We observe that when the weighting becomes imbalanced (e.g., $\beta_z$ is much lower relative to the other parameters), a degradation in performance occurs. 
Within this balanced range, our model remains stable across different combinations of these hyperparameters.

\section{AQ-SDR Details}
\label{appendix:dataset}
\begin{figure}[h]
    \centering
    \includegraphics[width=0.95\columnwidth]{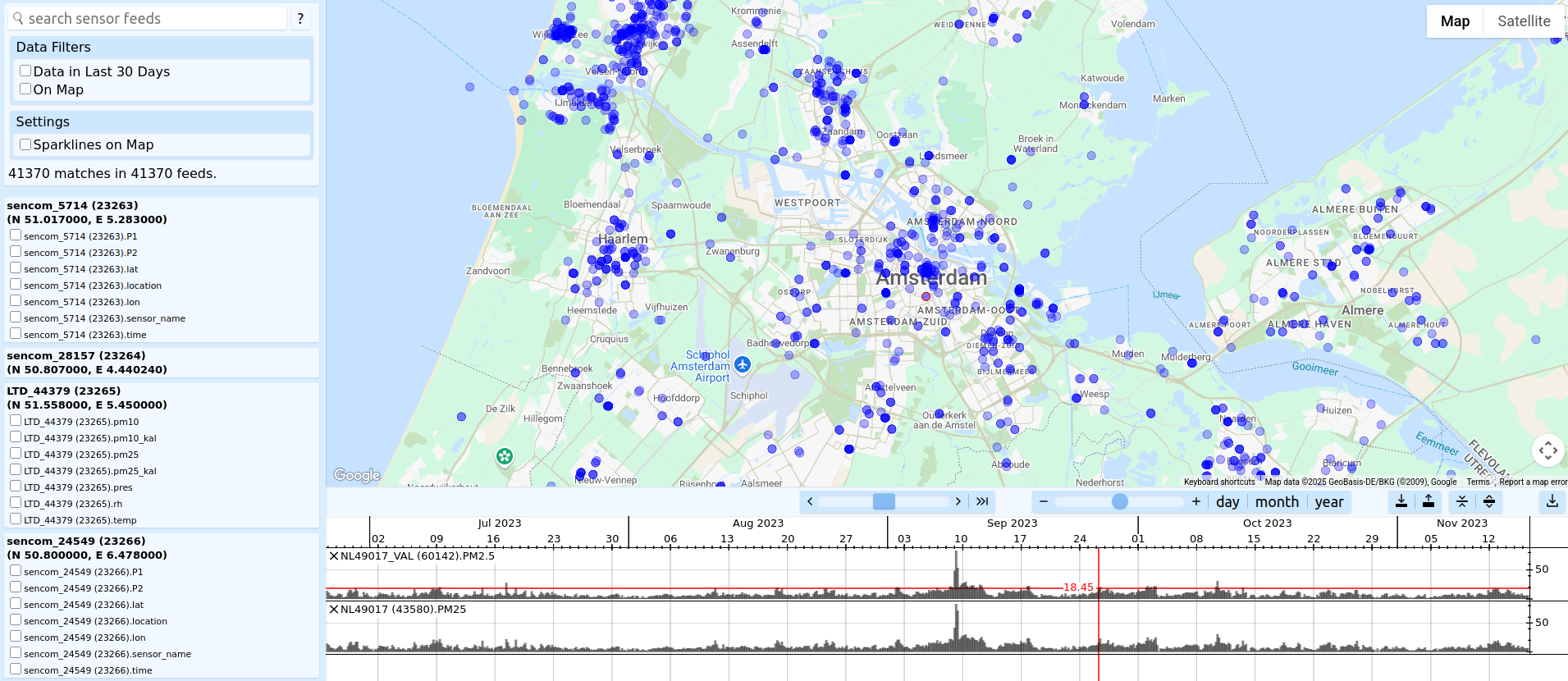} 
    \caption{A snapshot from the AQ-SDR interactive dashboard of sensors in the city of Amsterdam in the Netherlands. The report area at the bottom shows data streams from selected sensors in a chosen area near the city center.}
  \label{fig:dashboard_image}
\end{figure}

In this work, we introduced AQ-SDR, a large-scale collection of LCS measurements from Europe and Asia. 
Its purpose is to establish a common, standardized benchmark for developing and evaluating LCS calibration and correction techniques. An accompanying interactive dashboard visualizes the sensor distribution, as shown in Figure \ref{fig:dashboard_image}.

\subsection{Dataset Building}
\label{subsection:datasetbuilding}

AQ-SDR aggregates the LCS data from three major citizen-science initiatives: SamenMeten, Sensor.Community, and Location Aware Sensing System (LASS) community \cite{taiwan_data}. 
SamenMeten (Measure Together) is a Dutch citizen-science platform that supports public participation in environmental monitoring, focusing on air, water, and noise quality. 
Sensor.Community is a global citizen-science network that generates open AQ and noise data via LCS. Lastly, LASS is a Taiwanese citizen-science platform that leverages citizen-contributed AQ data and environmental sensing networks.

To supply reference measurements for supervised tasks or for validation purposes, we provide data from four authoritative sources: 1) LuchtMeetNet (Air Measurement Network), a Dutch AQ monitoring network that displays pollutant levels from maintained reference stations in real time; 2) the Royal Netherlands Meteorological Institute (KNMI), the Dutch national weather service and research institute for meteorology, climate, AQ, and seismology, which provides meteorological and environmental monitoring data; 3) the European Environment Agency (EEA), which provides independent environmental data and data platforms, and publishes regular reports; and 4) Taiwanese Ministry of Environment Open Data, which provides real-time streams from multiple high-precision reference stations across the country.

While these networks offer high-accuracy readings, their spatial coverage is sparse, preventing comprehensive reporting of AQ as noted earlier. 
The majority of the deployed LCS began operating in 2019 and are continually providing measurements, with a smaller subset being operational prior to 2019. 

\subsection{Preprocessing}

\label{subsection:preprocessing}

The dataset and the code to identically replicate the preprocessing will be made publicly available. 
First, we resample every sensor's data using its hourly average, since different devices provide readings at different temporal frequencies. 
We then define a range-validation step that applies hard bounds for all values for each pollutant and meteorological measurement. For example, we drop outdoor temperature readings outside the range [-50$^{\circ}$C, 70$^{\circ}$C], as they are implausible in an inhabited area.
Next, we split every data stream into batches of two months and pass it through a density-based outlier detection (DBSCAN) with a lenient threshold \cite{dbscan}. 
The purpose of this outlier detection is to eliminate long periods of abnormal readings.
Outliers indicating potential sensor failure are set to `NA' (e.g., an extended period of PM$_{2.5}$ values near 600 $\mu\mathrm{g}/\mathrm{m}^3$).
After this preprocessing step, we ensure that every sensor was functional (non-NA hours) for at least 6,000 hours (65\% of a full year), though not necessarily consecutive. 
This step prevents data streams from being empty or overly sparse.
The code and further details on preprocessing are documented and will be shared along with the dataset.

\bibliography{Veli}

@online{WHO2025HealthRisks,
  author   = {{World Health Organization}},
  title    = {Health risks},
  year     = {2018},
  url      = {https://www.who.int/teams/environment-climate-change-and-health/air-quality-energy-and-health/sectoral-interventions/ambient-air-pollution/health-risks},
  urldate  = {2025-05-09},
}

@article{Zhang2025PM2_5CVD,
  author       = {Zhang, Zhihang and An, Ran and Guo, Haoyan and Yang, Xuanru},
  title        = {Effects of PM2.5 exposure and air temperature on risk of cardiovascular disease: evidence from a prospective cohort study},
  journal      = {Frontiers in Public Health},
  volume       = {12},
  year         = {2024},
  article-number = {1487034},
  doi          = {10.3389/fpubh.2024.1487034},
  url          = {https://doi.org/10.3389/fpubh.2024.1487034},
  urldate      = {2025-05-09},
}

@article{zaini_pm25_2022,
	title = {{PM2}.5 forecasting for an urban area based on deep learning and decomposition method},
	volume = {12},
	issn = {2045-2322},
	url = {https://doi.org/10.1038/s41598-022-21769-1},
	doi = {10.1038/s41598-022-21769-1},
	number = {1},
	journal = {Scientific Reports},
	author = {Zaini, Nur’atiah and Ean, Lee Woen and Ahmed, Ali Najah and Abdul Malek, Marlinda and Chow, Ming Fai},
	month = oct,
	year = {2022},
	pages = {17565},
}

@article{Delaine2019Calibration,
  author    = {Delaine, F. and Lebental, B. and Rivano, H.},
  title     = {In Situ Calibration Algorithms for Environmental Sensor Networks: A Review},
  journal   = {IEEE Sensors Journal},
  year      = {2019},
  volume    = {19},
  number    = {15},
  pages     = {5968--5978},
  doi       = {10.1109/JSEN.2019.2910317},
  url       = {https://ieeexplore.ieee.org/document/8686160/},
  urldate   = {2025-05-09},
}

@article{Hagan2018Sensors,
  author    = {Hagan, David H. and Isaacman-VanWertz, Gabriel and Franklin, Jonathan P. and Wallace, Lisa M.\;M. and Kocar, Benjamin D. and Heald, Colette L. and Kroll, Jesse H.},
  title     = {Calibration and assessment of electrochemical air quality sensors by co-location with regulatory-grade instruments},
  journal   = {Atmospheric Measurement Techniques},
  year      = {2018},
  volume    = {11},
  number    = {1},
  pages     = {315--328},
  doi       = {10.5194/amt-11-315-2018},
  url       = {https://www.atmospheric-measurement-techniques.net/11/315/2018/amt-11-315-2018.html},
  urldate   = {2025-05-09},
}

@inproceedings{Maag2016PreDeployment,
  author    = {Maag, Balz and Saukh, Olga and Hasenfratz, David and Thiele, Lothar},
  title     = {Pre-Deployment Testing, Augmentation and Calibration of Cross-Sensitive Sensors},
  booktitle = {Proceedings of the 2016 European Conference on Wireless Sensor Networks (EWSN)},
  year      = {2016},
  pages     = {169--180},
  publisher = {Junction Publishing, Canada / ACM},
  doi       = {10.5555/2893711.2893735},
  url       = {https://doi.org/10.5555/2893711.2893735},
  urldate   = {2025-05-09},
}

@article{Concas2021LowCost,
  author       = {Concas, Francesco and Mineraud, Julien and Lagerspetz, Eemil and Varjonen, Samu and Liu, Xiaoli and Puolam{\"a}ki, Kai and Nurmi, Petteri and Tarkoma, Sasu},
  title        = {Low-Cost Outdoor Air Quality Monitoring and Sensor Calibration: A Survey and Critical Analysis},
  journal      = {ACM Transactions on Sensor Networks},
  volume       = {17},
  number       = {2},
  pages        = {1--44},
  year         = {2021},
  month        = may,
  doi          = {10.1145/3446005},
  url          = {https://doi.org/10.1145/3446005},
  urldate      = {2025-05-09},
}

@inproceedings{Cheng2020MapTransfer,
  author       = {Cheng, Yun and He, Xiaoxi and Zhou, Zimu and Thiele, Lothar},
  title        = {MapTransfer: Urban Air Quality Map Generation for Downscaled Sensor Deployments},
  booktitle    = {Proceedings of the 2020 IEEE/ACM Fifth International Conference on Internet-of-Things Design and Implementation (IoTDI ’20)},
  year         = {2020},
  pages        = {14--26},
  address      = {Sydney, Australia},
  month        = apr,
  publisher    = {IEEE/ACM},
  doi          = {10.1109/IoTDI49375.2020.00010},
  url          = {https://doi.org/10.1109/IoTDI49375.2020.00010},
  urldate      = {2025-05-09},
}

@inproceedings{Wang2023CaliFormer,
  author       = {Wang, Haoyang and Liu, Yuxuan and Zhao, Chenyu and He, Jiayou and Ding, Wenbo and Chen, Xinlei},
  title        = {CaliFormer: Leveraging Unlabeled Measurements to Calibrate Sensors with Self-Supervised Learning},
  booktitle    = {Adjunct Proceedings of the 2023 ACM International Joint Conference on Pervasive and Ubiquitous Computing \& the 2023 ACM International Symposium on Wearable Computing (UbiComp/ISWC ’23 Adjunct)},
  year         = {2023},
  pages        = {743--748},
  address      = {Cancún, Quintana Roo, Mexico},
  month        = oct,
  publisher    = {ACM},
  doi          = {10.1145/3594739.3612917},  
  url          = {https://doi.org/10.1145/3594739.3612917},
  urldate      = {2025-05-09},
}

@article{Cheng2019ICT,
  author         = {Cheng, Yun and He, Xiaoxi and Zhou, Zimu and Thiele, Lothar},
  title          = {ICT: In-field Calibration Transfer for Air Quality Sensor Deployments},
  journal        = {Proceedings of the ACM on Interactive, Mobile, Wearable and Ubiquitous Technologies},
  volume         = {3},
  number         = {1},
  article-number = {6},
  year           = {2019},
  month          = mar,
  doi            = {10.1145/3314393},
  url            = {https://doi.org/10.1145/3314393},
  urldate        = {2025-05-09},
}

@inproceedings{Wang2025AirRadar,
  author       = {Wang, Qiongyan and Xia, Yutong and Zhong, Siru and Li, Weichuang and Wu, Yuankai and Cheng, Shifen and Zhang, Junbo and Zheng, Yu and Liang, Yuxuan},
  title        = {AirRadar: Inferring Nationwide Air Quality in China with Deep Neural Networks},
  booktitle    = {Proceedings of the Thirty-Ninth AAAI Conference on Artificial Intelligence},
  volume       = {39},
  number       = {27},
  pages        = {28467--28475},
  year         = {2025},
  month        = apr,
  doi          = {10.1609/aaai.v39i27.35069},
  url          = {https://doi.org/10.1609/aaai.v39i27.35069},
  urldate      = {2025-05-09},
}

@inproceedings{rch2020,
author = {Li, Guodong and Ma, Rui and Liu, Xinyu and Wang, Yue and Zhang, Lin},
title = {RCH: robust calibration based on historical data for low-cost air quality sensor deployments},
year = {2020},
isbn = {9781450380768},
publisher = {Association for Computing Machinery},
address = {New York, NY, USA},
url = {https://doi.org/10.1145/3410530.3414322},
doi = {10.1145/3410530.3414322},
booktitle = {Adjunct Proceedings of the 2020 ACM International Joint Conference on Pervasive and Ubiquitous Computing and Proceedings of the 2020 ACM International Symposium on Wearable Computers},
pages = {650–656},
numpages = {7},
location = {Virtual Event, Mexico},
series = {UbiComp/ISWC '20 Adjunct}
}

@inproceedings{maximizingcorrelation,
author = {Li, Guodong and Wu, Zhiyuan and Liu, Ning and Liu, Xinyu and Wang, Yue and Zhang, Lin},
title = {Blind Calibration by Maximizing Correlation},
year = {2021},
isbn = {9781450384612},
publisher = {Association for Computing Machinery},
address = {New York, NY, USA},
url = {https://doi.org/10.1145/3460418.3480402},
doi = {10.1145/3460418.3480402},
booktitle = {Adjunct Proceedings of the 2021 ACM International Joint Conference on Pervasive and Ubiquitous Computing and Proceedings of the 2021 ACM International Symposium on Wearable Computers},
pages = {637–642},
numpages = {6},
location = {Virtual, USA},
series = {UbiComp/ISWC '21 Adjunct}
}

@ARTICLE{variationalbias,
  author={Li, Guodong and Wu, Zhiyuan and Liu, Ning and Liu, Xinyu and Wang, Yue and Zhang, Lin},
  journal={IEEE Sensors Journal}, 
  title={A Variational Bayesian Blind Calibration Approach for Air Quality Sensor Deployments}, 
  year={2023},
  volume={23},
  number={7},
  pages={7129-7141},
  doi={10.1109/JSEN.2022.3212009}}

@ARTICLE{survey2,
  author={Maag, Balz and Zhou, Zimu and Thiele, Lothar},
  journal={IEEE Internet of Things Journal}, 
  title={A Survey on Sensor Calibration in Air Pollution Monitoring Deployments}, 
  year={2018},
  volume={5},
  number={6},
  pages={4857-4870},
  doi={10.1109/JIOT.2018.2853660}}

@article{Jiao2016CAIRSENSE,
  title        = {Community Air Sensor Network (CAIRSENSE) project: Evaluation of low-cost sensor performance in a suburban environment in the southeastern United States},
  author       = {Jiao, Wan and Hagler, Gayle S.~W. and Williams, Ronald and Sharpe, Robert and Brown, Ryan and Garver, Daniel and Judge, Robert and Caudill, Motria and Rickard, Joshua and Davis, Michael and Weinstock, Lewis and Zimmer-Dauphinee, Susan and Buckley, Ken},
  journal      = {Atmospheric Measurement Techniques},
  volume       = {9},
  number       = {11},
  pages        = {5281--5292},
  year         = {2016},
  publisher    = {Copernicus Publications},
  doi          = {10.5194/amt-9-5281-2016}
}

@article{Diez2024QUANT,
  title        = {QUANT: a long-term multi-city commercial air sensor dataset for performance evaluation},
  author       = {Diez, Sebastian and Lacy, Stuart and Urquiza, Josefina and Edwards, Pete},
  journal      = {Scientific Data},
  volume       = {11},
  number       = {1},
  pages        = {904},
  year         = {2024},
  publisher    = {Springer Nature},
  doi          = {10.1038/s41597-024-03767-2}
}

@article{Bi2022,
  author  = {Bi, Jianzhao and Carmona, Nancy and Blanco, Magali N. and Gassett, Amanda J. and Seto, Edmund and Szpiro, Adam A. and Larson, Timothy V. and Sampson, Paul D. and Kaufman, Joel D. and Sheppard, Lianne},
  title   = {Publicly available low-cost sensor measurements for PM exposure modeling: Guidance for monitor deployment and data selection},
  journal = {Environment International},
  year    = {2022},
  volume  = {158},
  pages   = {106897},
  doi     = {10.1016/j.envint.2021.106897},
  url     = {https://doi.org/10.1016/j.envint.2021.106897}
}

@inproceedings{dbscan,
author = {Ester, Martin and Kriegel, Hans-Peter and Sander, J\"{o}rg and Xu, Xiaowei},
title = {A density-based algorithm for discovering clusters in large spatial databases with noise},
year = {1996},
publisher = {AAAI Press},
booktitle = {Proceedings of the Second International Conference on Knowledge Discovery and Data Mining},
pages = {226–231},
numpages = {6},
location = {Portland, Oregon},
series = {KDD'96}
}

@article{lcs_outdoor_survey,
author = {Concas, Francesco and Mineraud, Julien and Lagerspetz, Eemil and Varjonen, Samu and Liu, Xiaoli and Puolam\"{a}ki, Kai and Nurmi, Petteri and Tarkoma, Sasu},
title = {Low-Cost Outdoor Air Quality Monitoring and Sensor Calibration: A Survey and Critical Analysis},
year = {2021},
issue_date = {May 2021},
publisher = {Association for Computing Machinery},
address = {New York, NY, USA},
volume = {17},
number = {2},
issn = {1550-4859},
url = {https://doi.org/10.1145/3446005},
doi = {10.1145/3446005},
journal = {ACM Trans. Sen. Netw.},
month = may,
articleno = {20},
numpages = {44}
}

@inproceedings{
betavae,
title={beta-{VAE}: Learning Basic Visual Concepts with a Constrained Variational Framework},
author={Irina Higgins and Loic Matthey and Arka Pal and Christopher Burgess and Xavier Glorot and Matthew Botvinick and Shakir Mohamed and Alexander Lerchner},
booktitle={International Conference on Learning Representations},
year={2017},
url={https://openreview.net/forum?id=Sy2fzU9gl}
}

@ARTICLE{taiwan_data,
  author={Chen, Ling-Jyh and Ho, Yao-Hua and Lee, Hu-Cheng and Wu, Hsuan-Cho and Liu, Hao-Min and Hsieh, Hsin-Hung and Huang, Yu-Te and Lung, Shih-Chun Candice},
  journal={IEEE Access}, 
  title={An Open Framework for Participatory PM2.5 Monitoring in Smart Cities}, 
  year={2017},
  volume={5},
  number={},
  pages={14441-14454},
  doi={10.1109/ACCESS.2017.2723919}}

@inproceedings{lipschitz,
 author = {Virmaux, Aladin and Scaman, Kevin},
 booktitle = {Advances in Neural Information Processing Systems},
 editor = {S. Bengio and H. Wallach and H. Larochelle and K. Grauman and N. Cesa-Bianchi and R. Garnett},
 pages = {},
 publisher = {Curran Associates, Inc.},
 title = {Lipschitz regularity of deep neural networks: analysis and efficient estimation},
 url = {https://proceedings.neurips.cc/paper_files/paper/2018/file/d54e99a6c03704e95e6965532dec148b-Paper.pdf},
 volume = {31},
 year = {2018}
}

@InProceedings{ivae,
  title = 	 {Variational Autoencoders and Nonlinear ICA: A Unifying Framework},
  author =       {Khemakhem, Ilyes and Kingma, Diederik and Monti, Ricardo and Hyvarinen, Aapo},
  booktitle = 	 {Proceedings of the Twenty Third International Conference on Artificial Intelligence and Statistics},
  pages = 	 {2207--2217},
  year = 	 {2020},
  editor = 	 {Chiappa, Silvia and Calandra, Roberto},
  volume = 	 {108},
  series = 	 {Proceedings of Machine Learning Research},
  month = 	 {26--28 Aug},
  publisher =    {PMLR},
  url = 	 {https://proceedings.mlr.press/v108/khemakhem20a.html}
}

@inproceedings{vae,
  author = {Kingma, Diederik P. and Welling, Max},
  booktitle = {2nd International Conference on Learning Representations, {ICLR} 2014, Banff, AB, Canada, April 14-16, 2014, Conference Track Proceedings},
  eprint = {http://arxiv.org/abs/1312.6114v10},
  eprintclass = {stat.ML},
  eprinttype = {arXiv},
  title = {{Auto-Encoding Variational Bayes}},
  year = 2014
}

@article{lcs_variance,
title = {Performance and Applicability of Low-Cost PM Sensors to assess Global Pollution Variability through Machine Learning Techniques},
journal = {Atmospheric Environment: X},
pages = {100331},
year = {2025},
issn = {2590-1621},
doi = {https://doi.org/10.1016/j.aeaoa.2025.100331},
url = {https://www.sciencedirect.com/science/article/pii/S2590162125000218},
author = {Rajat Sharma and Andry Razakamanantsoa and Ashutosh Kumar and Thaseem Thajudeen and Agnès Jullien}
}

@Article{lcs_gaussian,
AUTHOR = {Malings, C. and Tanzer, R. and Hauryliuk, A. and Kumar, S. P. N. and Zimmerman, N. and Kara, L. B. and Presto, A. A. and R. Subramanian},
TITLE = {Development of a general calibration model and long-term performance
evaluation of low-cost sensors for air pollutant gas monitoring},
JOURNAL = {Atmospheric Measurement Techniques},
VOLUME = {12},
YEAR = {2019},
NUMBER = {2},
PAGES = {903--920},
URL = {https://amt.copernicus.org/articles/12/903/2019/},
DOI = {10.5194/amt-12-903-2019}
}

@inproceedings{pca,
 author = {Weston, Jason and Sch\"{o}lkopf, Bernhard and Bakir, G\"{o}khan},
 booktitle = {Advances in Neural Information Processing Systems},
 editor = {S. Thrun and L. Saul and B. Sch\"{o}lkopf},
 pages = {},
 publisher = {MIT Press},
 title = {Learning to Find Pre-Images},
 url = {https://proceedings.neurips.cc/paper_files/paper/2003/file/ac1ad983e08ad3304a97e147f522747e-Paper.pdf},
 volume = {16},
 year = {2003}
}

@inproceedings{
liu2024itransformer,
title={iTransformer: Inverted Transformers Are Effective for Time Series Forecasting},
author={Yong Liu and Tengge Hu and Haoran Zhang and Haixu Wu and Shiyu Wang and Lintao Ma and Mingsheng Long},
booktitle={The Twelfth International Conference on Learning Representations},
year={2024},
url={https://openreview.net/forum?id=JePfAI8fah}
}

@article{tesla,
      title={Real-Time Calibration Model for Low-Cost Sensor in Fine-Grained Time Series},
      author={Ahn, Seokho and Kim, Hyungjin and Shin, Sungbok and Seo, Young-Duk},
      journal={Proceedings of the AAAI Conference on Artificial Intelligence},
      year={2025},
      month={Apr.},
      pages={3-11},
      volume={39},
      number={1},
      url={https://ojs.aaai.org/index.php/AAAI/article/view/31974},
      DOI={10.1609/aaai.v39i1.31974},
}

@article{kalman,
    Author = {Kalman, Rudolph Emil},
    Title = {A New Approach to Linear Filtering and Prediction Problems},
    Journal = {Transactions of the ASME--Journal of Basic Engineering},
    Volume = {82},
    Number = {Series D},
    Pages = {35--45},
    Year = {1960}
}

@article{VanPoppel2023SensEURCity,
  title        = {SensEURCity: A multi-city air quality dataset collected for 2020/2021 using open low-cost sensor systems},
  author       = {Van Poppel, Martine and Schneider, Philipp and Peters, Jan and Yatkin, Sinan and Gerboles, Michel and Matheeussen, Christina and Bartonova, Alena and Davila, Silvije and Signorini, Marco and Vogt, Matthias and Dauge, Franck Ren and Skaar, Jran Solnes and Haugen, Rolf},
  journal      = {Scientific Data},
  volume       = {10},
  number       = {1},
  pages        = {322},
  year         = {2023},
  publisher    = {Springer Nature},
  doi          = {10.1038/s41597-023-02135-w}
}

\end{document}